\documentclass[twocolumn,prb]{revtex4-2}

\usepackage{amsmath,amssymb} 
\usepackage{amsmath} 
\usepackage{amsfonts} 
\usepackage{bm} 
\usepackage[colorlinks=true,citecolor=blue,linkcolor=blue,filecolor=blue,urlcolor=blue]{hyperref} 
\usepackage[dvipdfmx]{graphicx} 
\usepackage{braket} 

\usepackage{xcolor}
\usepackage{ulem}
\newcommand{\sh}[1]{{\color{blue}{#1}}}

\begin{document}

\title{
Environment-matrix-product operator for boundary-free large-scale quantum many-body simulations
}
\author{Souta Shimozono}
\email{shimozono-sota631@g.ecc.u-tokyo.ac.jp}
\author{Chisa Hotta}
\email{chisa@phys.c.u-tokyo.ac.jp}
\affiliation{Department of Basic Science, University of Tokyo, Meguro-ku, Tokyo 153-8902, Japan}
\date{\today}

\begin{abstract}
We propose an alternative to the infinite density-matrix renormalization approach for accessing quantum many-body states within a finite-size calculation that faithfully mimics the thermodynamic limit.
Our method constructs environment matrix product operators (MPOs) representing the Hamiltonian of semi-infinite regions surrounding 
the target system. 
Starting from the finite-size ground-state MPS, we contract its Hamiltonian representation to generate effective environment MPOs, which are then attached to a renewed finite system in a recursive manner.
This iterative embedding drives the system toward a bulk-like state with negligible finite-size effects.
The scheme requires no assumption of homogeneity and achieves unprecedentedly long real-time dynamics 
free from boundary reflections.
\end{abstract}

\maketitle
A central goal of condensed matter physics is to understand the intrinsic properties of materials 
that manifest in the collective phenomena in the bulk-size systems. 
To achieve such a description in theory or numerics, one must effectively reach the thermodynamic limit 
and remove the artificial influence due to finite size or from the system boundaries\cite{shibata2011}. 
In many methods, the thermodynamic limit is not possible to study directly 
and requires size scaling\cite{fisher1972,cardy1986,affleck1986}, 
while for most algorithms the computational cost increases exponentially with the system size. 
\par
In one and two-dimensional systems, there exist some tensor-network-based algorithms that overcome this limit by taking advantage of the invariance under translation in space. 
A prominent example is the infinite time-evolving block decimation (iTEBD) algorithm
\cite{vidal2007classical,orus2008infinite} 
for matrix product state (MPS)\cite{fannes1992finitely,klumper1993matrix,schollwock2011density},
which directly simulates the time evolution of infinite 1D spin chains in the thermodynamic limit. 
Its basic ideas have been incorporated into infinite matrix product state (iMPS) formalisms, 
that optimizes the unit matrix whose translational copies altogether will form a ground state of the infinite size system. 
Earlier infinite density matrix renormalization group (iDMRG) \cite{mcculloch2008infinite}, 
first proposed as part of the DMRG algorithms\cite{white1992density, white1993density}, 
turned out to give a close relationship to iTEBD, showing that the iterative process of 
inserting the sites into a finite chain will provide 
the translationally invariant MPS as a fixed-point. 
More recently, such an MPS is obtained using a variational scheme\cite{zauner2018variational}. 
\par
Nevertheless, for systems lacking strict translational invariance, such as those with impurities, 
physical boundaries, spatial modulations, or spatially non-uniform dynamics, one must still work with a finite subsystem. 
In such cases, constructing an appropriate environment that mimics the influence of 
the surrounding bulk becomes crucial. 
To obtain a bulk property within a finite-size treatment, 
we have previously made use of the scheme called sine-square deformation\cite{gendiar2011},  
that renormalizes the finite system such that the system center efficiently serves 
as a bulk state and the edges as environment\cite{hotta2012,hotta2013}, 
while the application is so far limited to static phases. 
In this work, we develop an MPS framework that embeds a finite system of size $N$ into 
a bulk-like environment, helping the subsystem to exhibit thermodynamic behavior 
while allowing for retaining local inhomogeneities. 
Such an algorithm can be applied to the calculation of quantum many-body dynamics after the quench, 
where the quasi-particles can pass through the boundaries into the environment, 
avoiding back reflections from the boundaries. 
\par
Previously, several approaches have been proposed to incorporate the extended role of boundaries 
into the matrix product state. 
A common strategy is to attach effective boundary tensors that represent the influence of 
the semi-infinite surroundings, as in the corner transfers matrix renormalization group (CTMRG)\cite{nishino1996corner,nishino1997corner} or boundary-MPS, which are applied to two-dimensional tensor network states. 
The infinite boundary condition was also proposed, 
which uses the transfer matrix of a pre-calculated iMPS ground state at the boundaries of the finite region
\cite{Phien2012infinite, phien2013dynamical, milsted2013variational,wu2020time}. 
This strategy, termed the "infinite-to-finite" approach, can work successfully 
in reproducing the dynamics of local excitations by reducing unphysical boundary reflections. 
However, when the system contains inhomogeneities, interfaces, or broken translational symmetry, 
the iMPS assumption can no longer be applied. 
\par
In this paper, we propose a contrasting ``finite-to-infinite'' approach, which we call the environment-embedding scheme. 
Instead of starting from an infinite system, we start by obtaining the ground state MPS wave function of finite size $N$. 
By projecting a Hamiltonian to the left and right halves of the ground state MPS and compressing the information, 
we obtain a set of environment MPO Hamiltonians, and embed them on both sides of the updated $N$-site system. 
After recursively performing this process, the $N$ site system attached to these environments 
exhibits physical properties that mimic those of $N\rightarrow\infty$, 
and further, successfully mitigates boundary reflections in dynamical simulations.

\begin{figure}
  \includegraphics[width=0.48\textwidth]{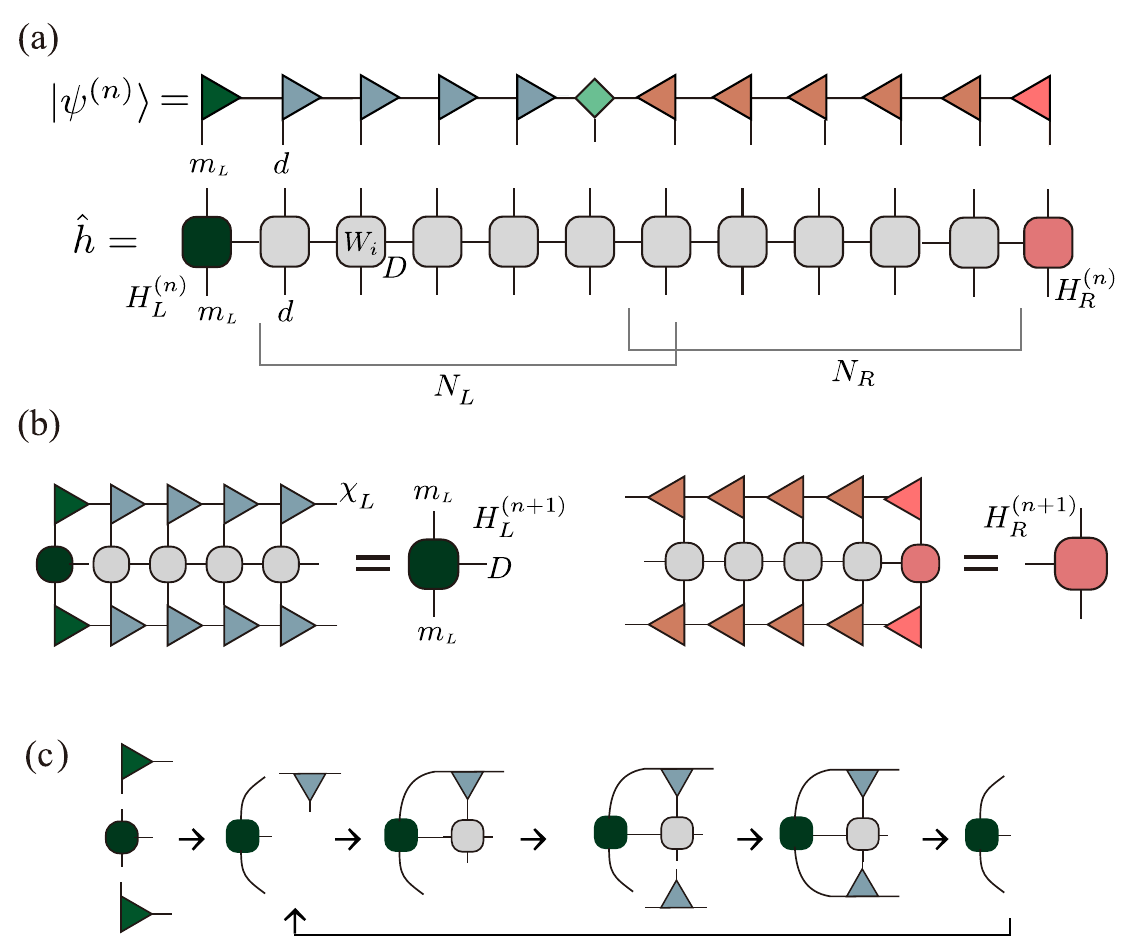}
  \caption{Graphical representation of (a) finite MPS with auxiliaries and MPO Hamiltonian, 
  $\hat h$, with embedded environments, $H_L$ and $H_R$. 
  (b) Matrix representation of the Hamiltonian (tensor) of the left subsystem of size $N_L$ in terms of 
      the Schmidt state $|\psi_L\rangle$, which becomes $H^{(n+1)}_L$ after the contraction. 
     Those of the right subsystem are also shown on the right. 
  (c) Process of contracting the tensor in (b), which is repeated by increasing $n_l$. }
  \label{f1}
\end{figure}
\textit{Finite MPS with environment tensors.} 
Let us consider an $N$-site MPS chain with environment sites with indices $L$ and $R$, 
which, later on, represent the semi-infinite system: 
\begin{align}
  |\psi\rangle = \sum_{s_L,\{s_i\},s_R} 
  A_L^{s_L}A_1^{s_1}\cdots A_N^{s_N} A_R^{s_R} |s_Ls_1 \cdots s_Ns_R\rangle.
\label{eq:gsmps}
\end{align}
For each site $i$, $\{s_i\}$ denotes a set of all physical degrees of freedom of dimension $d$, 
where $i=L,R$ contains larger degrees of freedom $m_{L/R}$, that represent the environments. 
Here, $A_i^{s_i}$ is a $\chi_{i-1}\times\chi_{i}\times d$ tensor and the environment tensors  
$A_L^{s_L}$ and $A_R^{s_R}$ are $m_{L/R}\times \chi$. 
Our goal is to obtain an energy eigenstate of the Hamiltonian of a bulk system 
given as a sum of local terms, e.g. nearest neighbor interactions or on-site field terms. 
The Hamiltonian is represented by a set of matrix product operators, 
$\{W_i\}_{i=1}^N$, of size $D\times D\times d \times d$, 
where $D$ is determined by the types of Hamiltonian. 
Assuming that a finite window of size $N$ is connected to the semi-infinite 
left and right environments,  
one can represent this MPO that acts on Eq.(\ref{eq:gsmps}) as 
\begin{equation}
\hat h= H_L \bigg(\prod_{i=1}^NW_i\bigg) H_R, 
\label{eq:hmpo}
\end{equation}
which is graphically represented in Fig.~\ref{f1}(a). A general MPO construction can be found in \cite{hubig2017generic}. 
For the transverse field Ising model(TFIM) with longitudinal field, 
${\cal H}=\sum_{i} -J\sigma_i^z\sigma_{i+1}^z - \Gamma \sigma^x_i + h \sigma^z_i$, 
we find 
\begin{align} 
& W = \begin{pmatrix}
    I & 0 & 0 \\
    -J\sigma^z & 0 & 0 \\
    -\Gamma\sigma^x +h\sigma^z & \sigma^z & I
  \end{pmatrix}, 
\label{eq:MPO_TFIM}
\end{align}
which we use as a concrete explanation in the following.
The lower triangular form allows us to write down the Hamiltonian of $N$ sites 
with an open boundary as, 
$\hat h^{(N)}= \begin{pmatrix}0& 0& 1\end{pmatrix}W^N \begin{pmatrix}1&0&0\end{pmatrix}^T$. 
\par
We first set the initial value as 
$H^{(0)}_L=(0, 0, 1)$ and $H^{(0)}_R=(1, 0, 0)^T$, 
and obtain the ground state $|\psi_0\rangle$, of $\hat h$ in the form, Eq.(\ref{eq:gsmps}), 
and transform it to the mixed canonical form shown in Fig.~\ref{f1}(b), 
$|\psi_0\rangle=\sum_{j=1}^{\chi_{i_c-1}} \lambda_j |\psi_l^{(j)}\rangle |\psi_r^{(j)}\rangle$, 
where the orthogonal center is chosen as site $i_c$, 
and $\{|\psi_{l/r}^{(j)} \rangle\}$ are the left/right Schmidt states with Schmidt values $\lambda_j$. 
We then prepare an MPO that represents the matrix element of the left $n_l=i_c-1$ site Hamiltonian, 
$\hat h_l=H_L\big(\prod_{i=1}^{n_l} W_i\big)$, given as 
\begin{align}
\left(H_l\right)_{i,j}= \langle \psi_l^{(i)}| \hat h_l |\psi_l^{(j)}\rangle
\label{eq:hl}
\end{align}
which has the dimension, $d^{n_l}\times d^{n_l} \times D$. 
This can be compressed one by one from the left as shown graphically in Fig.~\ref{f1}(c); 
first starting from $n_l=1$,
increasing $n_l$ by gradually shifting the orthogonal center from $i_c=2$ to $N_L$, 
truncating $d m_{l-1}$ to $m_l$ at each step.  
The resultant matrix that represents the $N_L$ sites is the left-environment MPO, $H_L^{(1)}$ 
of $m_L\times m_L\times D$. 
The same operation from the right edge will give the right-environment MPO, $H_R^{(1)}$, 
representing the rightmost $N_R$ sites. 
Then, we prepare another MPO consisting of $N$ center sites with left and right environments 
by plugging in $H_{R/L}^{(1)}$ to Eq.(\ref{eq:hmpo}). 
The subsequent iteration processes are given as follows: 
\begin{description}
  \item[1]{\bf Diagonalization.}
  Prepare an MPO Hamiltonian $\hat h$ with $\{W_n\}$, $H^{(n)}_R$ and $H^{(n)}_L$, 
  and obtain the MPS ground state $|\psi^{(n)}\rangle$ using the DMRG algorithm. 
  The initial choice of MPS tensors can be random matrices with a predefined bond dimension $\chi$. 
  \item[2]{\bf Left-contraction.}
  Performing a Schmidt decomposition of $|\psi^{(n)}\rangle$ by dividing the system into $n_l$ and $n_r$ sites, obtain a mixed canonical 
  form with $i_c=n_l+1$. Construct $H_l$ using Eq.(\ref{eq:hl}), and contract the bond dimension 
  from $\chi_l$ to $m_l$ within a given truncation error. Repeat this process by increasing $n_l$ from 2 up to $N_L=N/2$, 
  and obtain $H^{(n+1)}_L$, 
  \item[3]{\bf Right-contraction}
  Perform the same process for the right part of the system to obtain $H^{(n+1)}_R$. 
  \item[4]{\bf Embedding.}
  Update the MPO Hamiltonian by replacing the environment MPO with $H^{(n+1)}_L$ and $H^{(n+1)}_R$. 
  \item[5]
  Repeat steps from 1 to 4 until the state converges. 
\end{description}
There are two remarks; 
First, the rationale behind the choice of $N_L=N_R=N/2$ is to ensure that the junction of the environment 
is built primarily from the central part of the system, 
which is presumed to most accurately reflect the bulk property, 
or the least affected by the initial boundary conditions. 
Second, we need to notice the treatment to avoid numerical divergence. 
As we recursively contract the environment MPO, its major component that yields the energy density should grow 
linearly with the number of sites implicitly stored in the environment, $\sim nN/2$, 
that naturally leads to divergence. 
We thus introduce a traceless gauge that formally keeps the environment at a zero energy level, 
which, however, does not affect the functionality of the environment. 
In our construction on the TFIM in Eq.(\ref{eq:MPO_TFIM}), the $(3,1)$ element corresponds to the local environment 
energy term, and the condition to have its trace-zero is $\text{Tr}\big(H_L^{(n)} \begin{pmatrix}1 & 0 & 0\end{pmatrix}^T\big) = 0$. 
\begin{figure}[t]
  \includegraphics[width=0.48\textwidth]{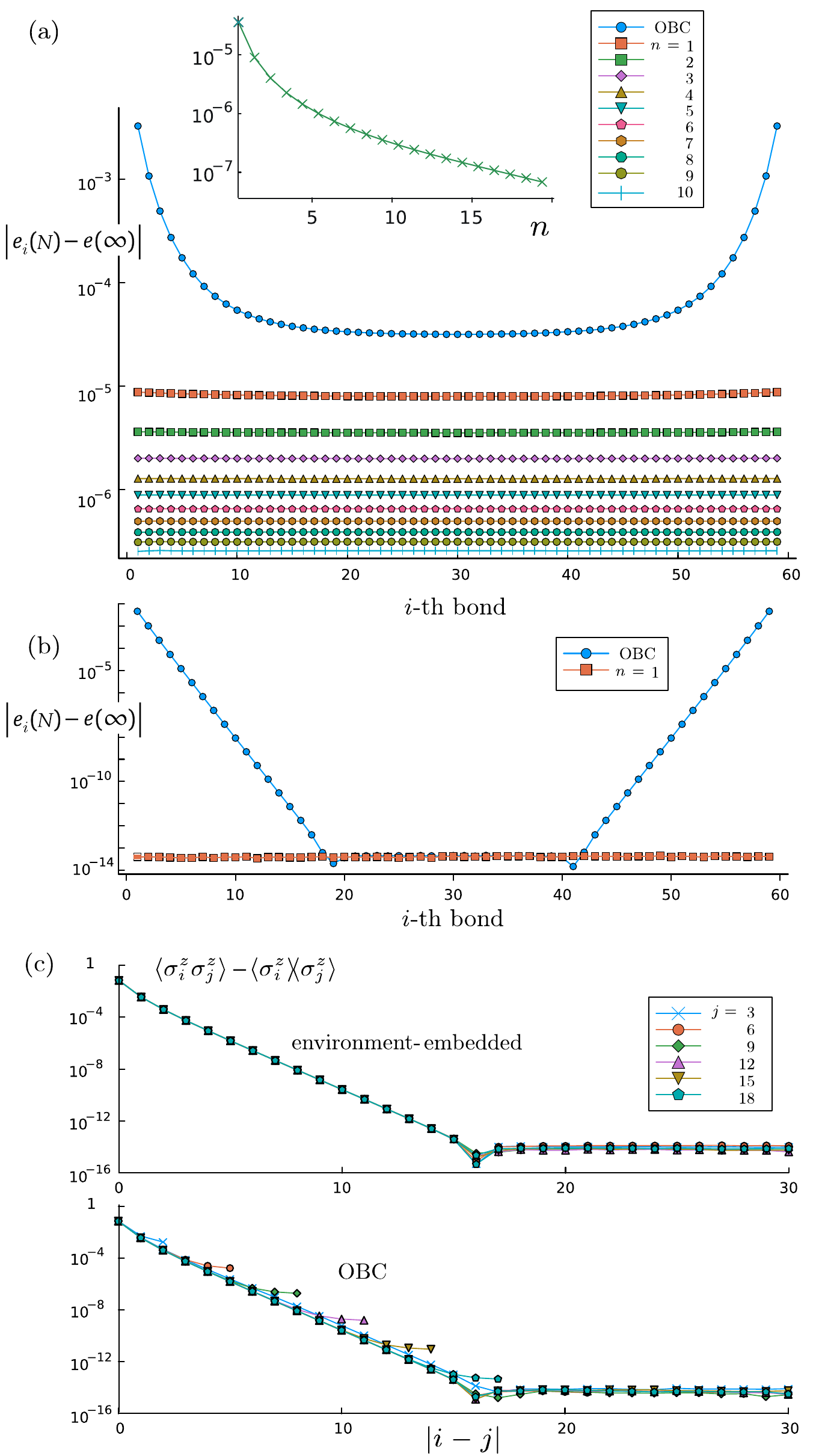}
  \caption{Static properties of the ground state of TFIM for a system of $N=60$ and 
  for the bond dimension, $m_{L/R}=30$, of the environment MPO. 
  (a) Energy density $e_i(N)$ at bond $i$ (with its both edge sites included by half) 
   for several choices of the number of embedding $n=1,2,\cdots,10$ 
   in the gapless critical point, $\Gamma/J=1$. 
   The $n$-dependence of $|e_i(N)-e(\infty)|$ at the center site for $\chi=30$ is shown in the inset. 
  (b) $e_i(N)$ for the gapped state $\Gamma/J=0.5$ with $n=1$, in comparison with OBC ($n=0$). 
  (c) Spin-spin correlation function 
  $\langle \sigma_i^z \sigma_j^z \rangle - \langle \sigma_i^z\rangle\langle\sigma_{j}^z \rangle$ with 
   fixed values of $j$, for the gapped phase $\Gamma/J=0.5$.  
   The top and bottom panels show the case of embedding $n=1$ and OBC, respectively. 
}
  \label{f2}
\end{figure}

\par
{\it Recovery of bulk translational symmetry.} 
We benchmark the efficiency of the environment embedding using the TFIM with zero longitudinal field $h=0$. 
In Fig.~\ref{f2}(a) we show the energy density $e_i(N)$ measured for the $i$-th bond 
after embedding the environment $n$ times at the gapless point, $\Gamma/J=1$, 
against $e(N\rightarrow \infty)$. 
For the OBC, there exists a large deviation of $e_i(N)$, which is a boundary effect, 
decaying at $\sim 1/L$. 
However, after embedding even once, $n=1$, one finds a perfect recovery of translational symmetry. 
The difference from $e(\infty)$ is suppressed much faster than the logarithm of $n$, 
as we find from the inset of the figure. 
\par 
For the gapped state, $\Gamma/J=0.5$, as shown in Fig.~\ref{f2}(b), 
the embedding immediately realizes the bulk energy 
with the translational symmetry. Due to the gap, the MPS works well, and even for OBC 
$e_i(N)$ is comparably accurate at the center one-third of the system. 
To strictly examine the recovery of bulk translational symmetry, we calculate the spin-spin correlation 
$\langle \sigma_i^z \sigma_j^z \rangle - \langle \sigma_i^z\rangle\langle\sigma_j^z \rangle$ in Fig.~\ref{f2}(c) 
as a function of $|i-j|$.  
This definition, subtracting the magnetization of the center site rather than the local one, 
serves as a sensitive probe for boundary induced deviations.
The exponential decay, indicating the spin gap, is clearly observed 
down to the value of $10^{-10}$, which is enough to judge that the numerical accuracy 
of translational invariance is precisely recovered. 
For OBC, consistent with the large deviation observed in $e_i(N)$, the correlation between the spins near the edge 
 ($j=3,6$) fails to exhibit the exponential decay and saturates at a finite value, indicating that near the edge 
the boundary effect is serious.  
\par
\begin{figure}[t]
  \includegraphics[width=0.48\textwidth]{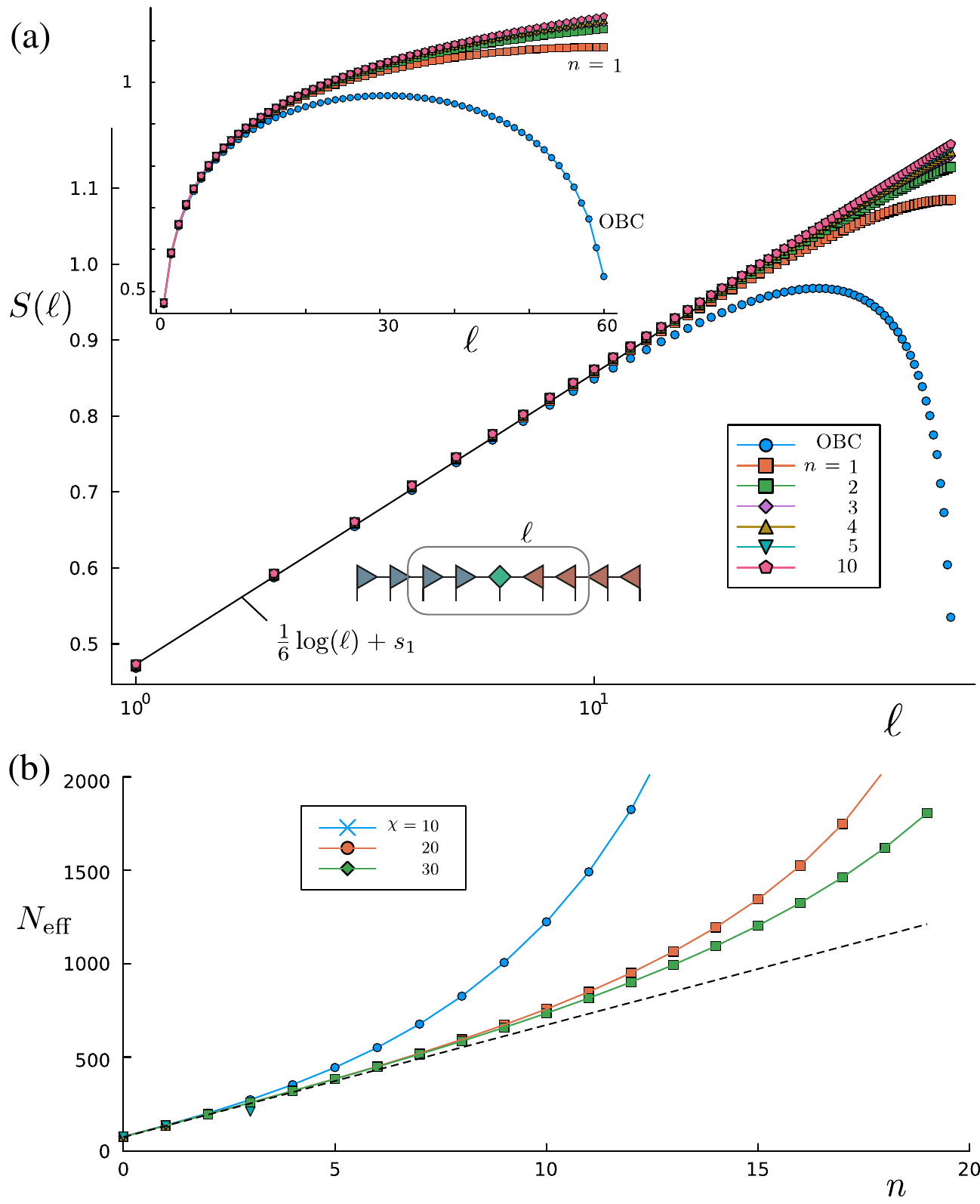}
  \caption{(a) Entanglement entropy $S(\ell)$ as a function of subsystem size $\ell$ 
  of the TFIM at $\Gamma/J=1$ and $N=60$ and $\chi=30$ for several choices of embedding, 
  $n=1,2,\cdots,5,10$, and for OBC. 
  The solid line in the semilog plot indicates the function $\frac{c}{3}\log(\ell)$ with $c=1/2$. 
  (b) Effective site length $N_{\rm eff}$, obtained by fitting the EE 
  when we assume the form $S_{\rm OBC}(\ell)$ with $c=1/2$ to the environment-embedded data, 
  where we compare the cases, $\chi=10,20,30$. 
}
  \label{f3}
\end{figure}

In order to examine whether the environment-embedded Hamiltonian 
correctly captures the critical properties in the thermodynamic limit, 
we compute the entanglement entropy (EE) \cite{calabrese2004entanglement}. 
Figure~\ref{f3}(a) shows the EE, $S(\ell)$ as a function of subsystem size $\ell$, 
where we cut out the center $\ell$ sites after embedding the environment $n=1,2,3,4,5,10$ times. 
We first focus on the EE of the OBC system given as a reference, 
showing a convex upward curve with a strong suppression at $\ell \sim 1$ and $\sim N$ that is known to follow 
$S_{\rm OBC}(\ell) = \frac{c}{3}\log\left(\frac{2N}{\pi a}\sin\left(\frac{\pi \ell}{N}\right)\right) + s_1$. 
By fitting the data, we obtain $c=0.53827$ for $\chi=30$, and may require 
further extrapolation with $\chi$ \cite{tagliacozzo2008scaling, pollmann2009theory}. 
\par
Whereas for the environment-embedded case, $S(\ell)$ curve extrapolates to the one expected for 
the thermodynamic limit, $S(\ell) = \frac{c}{3}\log(\ell) + s_1$, shown together in 
the figure as solid line for $c=1/2$, known for the critical TFIM. 
The extrapolation of the data to $S(\ell)$ using up to $n=20$ yields $c=0.49962$ for fixed $\chi=30$.  
Notably, the continuously increasing $S(\ell)$ up to the whole system size, $\ell=N$, 
strongly indicates that the system realizes a nearly $N=\infty$ state with the support from the environment. 
We plot in Fig.~\ref{f3}(b) the effective system size $N_{\rm eff}$ as a function of the number of embedding $n$, 
extracted from the $\ell$ dependence of EE 
when assuming the form $S_{\rm OBC}(\ell)$ with $c=1/2$. 
At $n=0$ (OBC), we start from $N_{\rm eff}=62=N+2$ (where an additional 2 sites come from the environment), 
and it increases rapidly with $n$. The value extrapolated by $\chi$ gives the linear curve 
that gives $N_{\rm eff}= nN+(N+2)$, which demonstrates the fact that each embedding step effectively adds $N$ sites to the environment. 
This result indicates that the environment tensor successfully captures the contracted system size. 
\par
%
\par
\noindent
{\it Dynamical properties.}
Using the environment-embedded state as a platform, we now calculate the real-time dynamics 
of the TFIM at $\Gamma=J=1$ with a small longitudinal field $h=0.05$ (TFIM+L). 
The low-energy excitations of this parameter range are described by 
the $E_8$ integrable quantum field theory \cite{zamolodchikov1989integrals}, 
where there appears a series of excitation modes around $k=0$, 
whose energy levels are established together with the analytical quasi-particle masses.  
Indeed, ever since such a spectrum was observed in the neutron diffraction experiment performed on 
quasi-1D Ising ferromagnet CoNb$_2$O$_6$ \cite{coldea2010quantum}, and theoretically studied \cite{kjall2011bound, woodland2023excitations}. Recently, $E_8$ particles were also observed in BaCo$_2$V$_2$O$_8$\cite{zhang2020observation,zou2021e8,Amelin2022}.
The TFIM+L was used to benchmark the newly developed numerical technique for real-time dynamics\sh{\cite{albert2024truncated}}. 
\par
\begin{figure*}
  \includegraphics[width=1.0\textwidth]{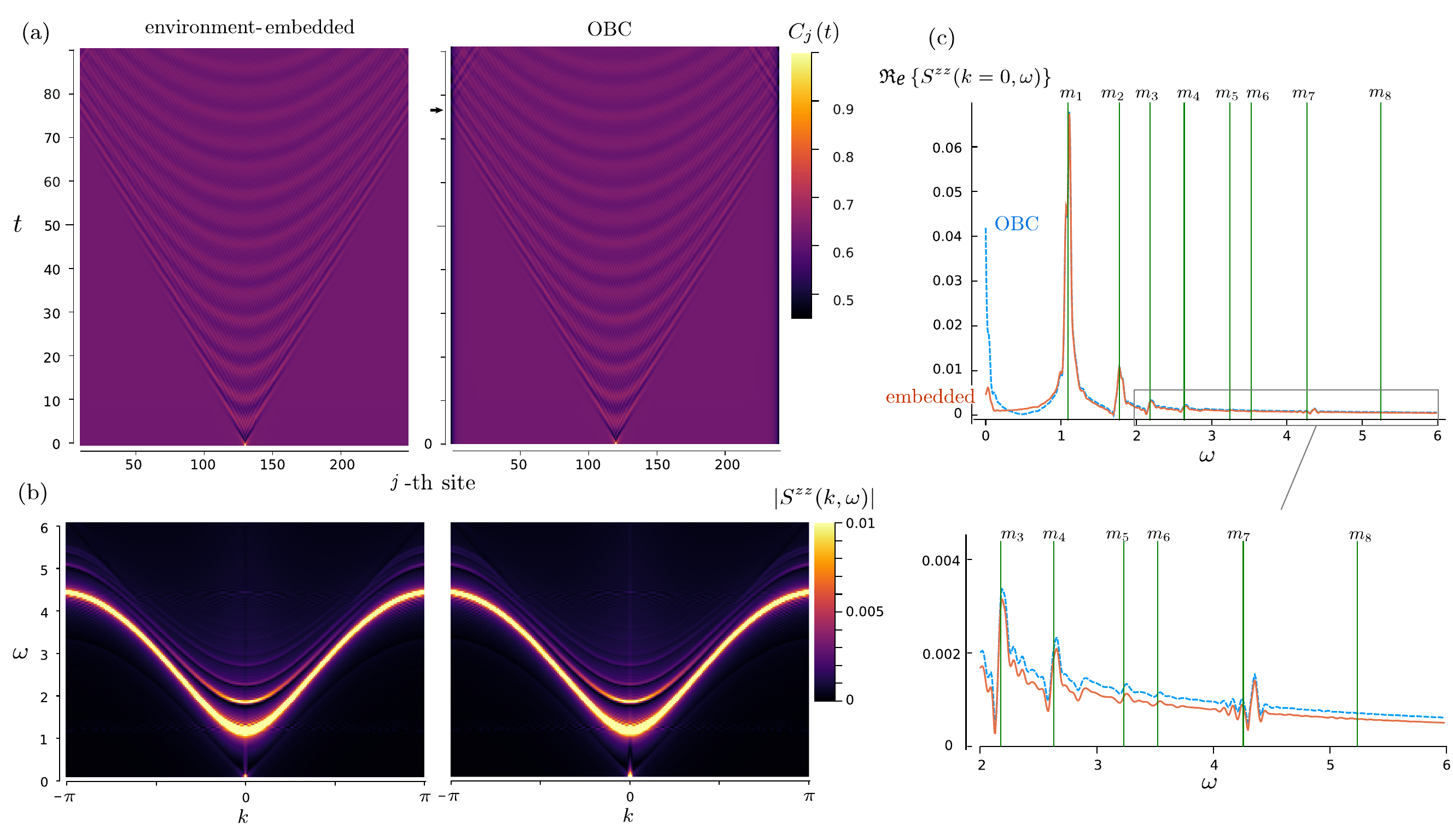}
  \caption{Real-time dynamics of TFIM+L in a $E_8$ regime with $\Gamma=J=1$, $h=0.05$. 
(a) Density plot of $C_j(t)=|\langle\sigma_j(t)\sigma_{N/2}(0)\rangle|$ for $N=240$ 
with $n=1$ embedding (left panel) and OBC (right). 
(b) Density plot of the dynamical structure factor $S^{zz}(k, \omega)$ for $N=240$. 
The Fourier transform of $C_j(t)$ uses Hann window to position space, 
and Gaussian window $\sigma=\sqrt{T^2/(-2*\log(\alpha))}$ to time space \cite{drescher2023dynamical}, 
with a linear extrapolation\cite{white2008spectral,barthel2009spectral}, using the data $t=[0:85]$.
(c) $\mathfrak{R}e\{S^{zz}(k=0, \omega)\}$ for the data in panel 
 (b) with embedding (solid line) and OBC(broken line). 
 Vertical lines denote the $E_8$ spectrum from the field theory. 
  }
  \label{f4}%
\end{figure*}

We first obtain the ground state $\ket{\psi_{0}}$ by the environment embedding, 
and quench the center spin at $t=0$ as $\ket{\phi(t=0)}=\sigma_{N/2}^z \ket{\psi_{0}}$, and perform the unitary time evolution, 
for which we used the two-site version of the time-dependent variational principle(2-TDVP)
\cite{haegeman2011time, haegeman2016unifying, paeckel2019time} 
with a timestep of $dt=0.04$ and measurement timestep of $\delta t=0.20$ 
with max bond dimension fixed to $\chi=100$. 
By the Fourier transform of the dynamical correlation function during the time evolution, 
we obtain the dynamical structure factor as 
\begin{align}
  S^{zz}(k, \omega) = \int dt \sum_{j} e^{i(\omega t - k j)} \langle \sigma_j^z(t) \sigma_{N/2}^z(0) \rangle. 
\end{align}
Figure~\ref{f4}(a) shows the absolute value of the dynamical correlation function 
$C_j(t)=|\langle \sigma_j^z(t) \sigma_{N/2}^z(0) \rangle|$, 
of the environment-embedded and the OBC systems at $N=240$. 
In OBC, the wave front of the excited quasi-particles develops and forms a light-cone 
and are reflected at the boundaries, which leads to the interference of waves. 
The advantage of using our environment embedding is to suppress such reflection. 
Indeed, for the OBC data there clearly exists a reflection at around $t\sim 75$ indicated by an arrow. 
Whereas, for the environment embedding, there is no signal from the boundaries until $t\lesssim 90$. 
\par
The difference between the two manifests at the lowest energy part of the spectrum in $S^{zz}(k, \omega)$; 
As shown in Fig.~\ref{f4}(b), 
the OBC spectrum has an intense artificial $\omega=0$ peak, which is more clearly observed for 
the $k=0$ slices in Fig.~\ref{f4}(c). 
For reference, we show the quasi-particle peak positions predicted from the $E_8$ field theory. 
Up to $m_7$, the peak positions are well reproduced for both the environment-embedded and the OBC, 
as these peaks are extracted from the oscillations up to the time of reflections, 
consistent with the previous work using TEBD\cite{kjall2011bound}. 
In Table I, we summarize the numerical values of six different mass ratios for $N=100, 160$, and $240$, 
in comparison with the $E_8$ field theory. 
\par
\par
Previously, there were infinite boundary conditions that are to be compared with the present work\cite{Phien2012infinite};  
they first obtain a translationally invariant MPS and construct a semi-infinite environmental MPS (not the MPO), 
attached to both sides of the size-$N$ system. 
When they apply it to the real-space dynamics, the wave front was seen to propagate beyond the infinite boundaries 
without back reflection. 
Their treatment appears to require strict control of the iMPS as the fixed point of the renormalization process, 
and has a prerequisite of targeting a uniform Hamiltonian. 
Regarding dynamical calculations, 
there is also another series of schemes to use iMPS in real space and develop it in the time-direction in the form of MPO 
about the physical quantities\cite{Banuls2009,Banuls2022}, which also relies on the spatial homogeneity to obtain the iMPS. 
In contrast, our ``finite-to-infinite'' scheme does not rely on pre-existing iMPS or the assumption of homogeneity, 
which, in principle, suggests a wider flexibility. 
It can be applied to spatially inhomogeneous Hamiltonians, or to more flexible conditions like  
quenching a wider range of the system whose influence immediately reaches the boundaries, 
or even perturbing the system during the dynamics. 
\par
\begin{table}[tbh]
\centering
\begin{tabular*}{\columnwidth}{@{\extracolsep{\fill}}lcrrrr}
 Ratio & $E_8$  & $N=100$ & $N=160$ & $N=240$ &   \\  
\toprule
$\omega_2/\omega_1$ & 1.618  & 1.648 & 1.593  &  1.628  \\
$\omega_3/\omega_1$ & 1.989  & 2.042 & 2.000  &  2.013  \\
$\omega_4/\omega_1$ & 2.405  & 2.451 & 2.395  &  2.445  \\
$\omega_5/\omega_1$ & 2.956  & 2.971 & 2.930  &  2.993  \\
$\omega_6/\omega_1$ & 3.218  & 3.239 & 3.162  &  3.256  \\ 
$\omega_7/\omega_1$ & 3.891  & 3.901 & 3.883  &  3.905  \\ 
\end{tabular*}
\caption{Peak ratio of the $E_8$ quasi-particle states obtained by analyzing the data in Fig.~\ref{f4}, 
in comparison with the $E_8$ Toda field theory.}
\label{tab:e8_ratio}
\end{table}
{\it Summary.} 
We proposed the environment-embedding scheme, 
which iteratively constructs an environmental Hamiltonian operator 
from the mixed canonical MPS eigenstate of the finite-size Hamiltonian. 
The basic strategy is to store major information of the Hamiltonian 
in terms of the ground state, 
by compressing the dimension of the Hamiltonian matrix to those providing the large Schmidt values. 
As we do not need to keep the information on the degrees of freedom of the environment, 
the compression can be done as many times as one wants with a minimal cost in memory. 
\par
In some aspects, our method shares a common concept with the density matrix embedding theory (DMET), 
designed for fermionic systems\cite{knizia2012,wouters2016}.
This method divides the large-size cluster into a small system and the environment. 
and projecting the original Hamiltonian to some particular Schmidt basis sets 
connecting the small system and the environment at the noninteracting level. 
The ground state of the resultant effective Hamiltonian exhibits almost identical 
entanglement spectrum between the small system and the environment\cite{plat2020,kawano2020}. 
Although the discipline to choose the basis to project is different from ours, 
both our method and theirs optimize the environment Hamiltonian. 
\par
The process of constructing the environment in our method 
resembles the early-period DMRG that uses a two-block plus the center two-site scheme 
(we call block-state based DMRG), 
where the information of the left and right blocks are compressed to a minimum dimension, 
are stored as a matrix element of the operator. 
However, as we frequently observe in the usual DMRG calculation\cite{shibata2011}, 
the ground state suffers serious effects from the boundaries, 
such as Friedel oscillations that only decay slowly with inverse distance from the boundaries.  
In that respect, the $N$-site system with $n=1$ embedding far better serve as a nearly bulk state, 
compared to the $N+N=2N$ states with OBC. 
The key to such success shall be that the center $N$ sites are optimized with 
the aid of the environment keep only the 
distilled major information (see Fig.~\ref{f3}(b), where smaller $\chi$ has larger $N_{\rm eff}$). 
\par
There is also an advantage against the block-state-based DMRG: 
while our environmental MPO works similarly to the block-state of DMRG, 
the Hamiltonian of the main $N$ site system is a product of a single-site MPO, 
and the wave function is controlled locally, 
making full advantage of the canonical forms of MPS that fix the gauges from both ends. 
This helps us to overcome the limitations of conventional finite-size methods or 
infinite-size methods that assume homogeneity, 
and attains a high tunability and applicability to various systems:  
e.g. those with random quenched interactions, where one can construct an environment for one 
types of randomness, and take an average over many random samples. 
The real-time dynamics do not have to rely on the local quench, and can simulate a fictitious 
nonuniform time-evolving process to meet their needs. 
The method itself is conceptual, and does not limit itself to the 1D MPS, but to the 
other techniques like PEPS in higher dimensions. 

\begin{acknowledgments}
This work is supported by a Grant-in-Aid for Transformative Research Areas 
``The Natural Laws of Extreme Universe---A New Paradigm for Spacetime and Matter from Quantum Information" 
(Grant No. 21H05191) and other JSPS KAKENHI (No. 21K03440). 
S.S. thanks Masataka Kawano, Atsushi Iwaki, and Hidehiro Saito for fruitful discussions. The numerical simulations were performed on the Yukawa-21 supercomputer at the Yukawa Institute for Theoretical Physics, Kyoto University. We acknowledge support from the World-leading-Innovative Graduate Study Program of the Advanced Basic Science Course, the University of Tokyo. All the calculations were performed using the ITensor library \cite{itensor}.
\end{acknowledgments}

\bibliography{envbib20251130}

\end{document}